\documentclass{aa}
\usepackage{graphicx}
\usepackage[varg]{txfonts}
\usepackage{natbib}
\bibpunct{(}{)}{;}{a}{}{,} 
\usepackage{multirow}
\usepackage{epsfig}
\usepackage[figuresright]{rotating}
\usepackage{natbib}
\bibpunct{(}{)}{;}{a}{}{,} 

\begin{document}

   \title{Processing of analogues of plume fallout in cold regions of Enceladus by energetic electrons}


   \author{A. Bergantini
          \inst{1}
          \and
          S. Pilling\inst{1} 
          \and
          B. G. Nair\inst{2}
          \and
          N.J. Mason\inst{2}
          \and
          H.J. Fraser\inst{2}
          }

   \institute{Universidade do Vale do Paraiba (UNIVAP), Instituto de Pesquisa e Desenvolvimento (IP\&D), Sao Jose dos Campos, SP, Brazil.\\
              \email{alebergantini@hotmail.com}
         \and
             Department of Physical Sciences, The Open University, Milton Keynes MK7 6AA, UK.\\
             }

   \date{Received /
            accepted }


  \abstract
   {Enceladus, a small icy moon of Saturn, is one of the most remarkable bodies in the solar system. This moon is a geologically active object, and despite the lower temperatures on most of its surface, the geothermally heated south polar region presents geysers that spouts a plume made of water ($\sim$90\%), carbon dioxide, methane, ammonia, and methanol, among other molecules.
   Most of the upward-moving particles do not have the velocity to escape from the gravitational influence of the moon and fall back to the surface. The molecules in the ice are continuously exposed to
   ionizing radiation, such as UV and X-rays photons, cosmic rays, and electrons. Over time, the ionizing radiation promotes molecular bond rupture, destroying and also forming molecules, radicals, and fragments.}
   {We work analyse the processing of an ice mixture analogue to the Enceladus fallout ice in cold resurfaced areas (north pole) by 1 keV electrons. The main goal is to search for complex species that have not been detected yet in to this moon, and determine relevant physico-chemical parameters, such as destruction and formation cross-sections and the half-life of the studied molecules in the ice.}
   {The experiment consisted of the electron irradiation of an Enceladus-like ice mixture (H$_2$O:CO$_2$:CH$_4$:NH$_3$:CH$_3$OH) in an ultra-high vacuum chamber at 20 K. The analysis was made by infrared spectrometry in the mid-infrared region (4000-800 cm$^{-1}$ or 2.5-12.5 $\mu$m).}
   {The absolute dissociation cross-sections of the parent molecules, the formation cross-section of daughter species, and the half-life of the parental species in a simulated Enceladus irradiation scenario were determined.
   Among the produced species, CO (carbon monoxide), OCN$^-$ (cyanate anion), HCONH$_2$ (formamide), and H$_2$CO (formaldehyde) were tentatively detected.}
   {}


   \maketitle
%

\section{Introduction}

Enceladus is a tiny, but extremely interesting moon of Saturn. It is the smallest satellite in the Solar system that is still geologically active (\citealt{karg2006}). Enceladus presents an abnormal temperature gradient, with a hot spot in the south pole, a region permeated by large canyons where a plume of gas and grains erupts from ridges in the ground at $\sim$~200kg s$^{-1}$ rate (\citealt{hans2006}; \citealt{hans2008}). This plume was collected and analysed in situ by Cassini' instruments (especially the Ion and Neutral Mass Spectrometer (INMS) and the Ultraviolet Imaging Spectrograph (UVIS)), and the composition of the plume was determined to consist of $\sim$90\% of H$_2$O, $\sim$5\% of CO$_2$, $\sim$0.9\% of CH$_4$, $\sim$0.8\% of NH$_3$, plus small amounts of other species, which include CH$_3$OH, N$_2$, $^{40}$Ar, C$_n$H$_n$ and C$_n$H$_n$O (\citealt{spen2006}; \citealt{spen2009}). The mean velocity with which this material is ejected from the fractures in Enceladus' south pole is not enough for most of the gas and grains to escape and supply Saturn's E-ring. Indeed, models made by \cite{porc2006} predict that $\sim$99\% of the upward-moving particles fall back to the surface, resulting in vast resurface process due to plume fallout (\citealt{karg2006}), as evidenced by the largely snow-fresh non-cratered regions on the surface of Enceladus. In this sense, we believe that the composition of the plume, as detected by the Cassini mission, represents of the actual composition of Enceladus' ice in resurfaced regions very well.

Molecules above the surface of Enceladus are subjected to all kinds of ionizing radiation, including electrons from the solar wind and from the magnetosphere of Saturn, which is heavily connected with Enceladus' north polar region (\citealt{jone2006}; \citealt{pryo2011}). This is the coldest region of Enceladus' surface, with temperatures of up to 33 K (\citealt{spen2006}). The bombardment of the Enceladus' plume and resurfaced ice by energetic electrons may induce changes in the molecules, destroying and producing species, thus leading an the increased chemical complexity in the environment. Our main objective is to analyse the possible formation of complex molecules by processing the relatively simple molecules detected in Enceladus by NASA's Cassini mission.

Our study consists in the bombarding of an ice mixture (H$_2$O:CO$_2$:CH$_4$:NH$_3$:CH$_3$OH) at 20~$\pm$~2 K by 1 keV electrons inside an ultra-high vacuum chamber. The 1 keV electrons were employed to mimic the effect of fast particles from Saturn's magnetosphere (\citealt{brow2006}) and from the solar wind over Enceladus' surface, and the energy transfer from processes that occur in the track of possible MeV cosmic-ray trajectories (\citealt{jone2011}). The irradiation temperature (20~$\pm$~2 K) is compatible with the coldest regions of the north pole (where the temperatures are around 35 K), tracking physicochemical changes in the resurfaced layers of ice on Enceladus.

The molecules of the ice mixture in the experiment were determined based on some of the most abundant elements detected by instruments onboard the Cassini mission. Following \cite{spen2009}, H$_2$O and CO$_2$ are the two most abundant molecules in Enceladus' plume). \cite{hans2008} ascribe the darkness of Enceladus' surface at far ultraviolet (FUV) wavelengths due to the presence of tholin, complex organic molecules that results from mixtures of CH$_4$, NH$_3$, and H$_2$O under the influence of ultraviolet photons. We added methanol to account for a complex organic component in the mixture. According to \cite{bern1995}, some of the products from the processing of water and ammonia-containing ices are completely different in presence of methanol.


\section{Experimental set-up}

\begin{figure}
      \centering
      \resizebox{\hsize}{!}{\includegraphics{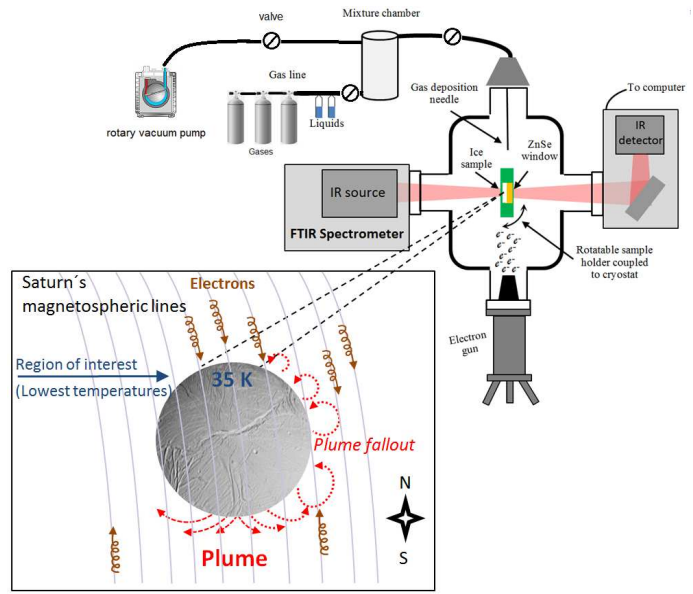}}
      \caption{Experimental apparatus at the Open University Astrochemistry Laboratory (not to scale), with the simulated region of Enceladus highlighted in the lower-left box.}
      \label{chamber}
  \end{figure}

The experiments to simulate analogues of plume fallout in some regions of Enceladus surface under the influence of energetic electrons were carried out in an ultra-high vacuum stainless steel chamber (base pressure below 5$\times$10$^{-9}$ mbar). The chamber was evacuated using an oil-free magnetically suspended turbo-molecular pump (Pfeiffer vacuum), coupled with a rotary pump. A closed-cycle helium cryostat (Sumitomo HC-4E) cooled down a freely rotatable cold finger to 20~$\pm$~2 K. The cold finger held a zinc selenide (ZnSe) crystal onto which the samples were deposited. The molecules were adsorbed in the cold finger after being mixed in a pre-chamber connected to an independent rotary pump vacuum system. The partial pressure of the gases admitted in the pre-chamber was measured individually. All the molecules used in the experiments were purchased commercially from the BOC Group and are research quality grade (minimum purity of 99.999\%). The H$_2$O and CH$_3$OH samples were prepared from the vapour pressure of liquid, purified through several freeze-thaw-pump cycles. The gas mixture was deposited through a capillary stainless tube held a few millimeters from the ZnSe target, for 20 minutes at a background pressure in the main chamber of 1$\times$10$^{-7}$ mbar. The experimental vacuum chamber is equipped with a freely rotatable sample holder coupled to a cryostat, operated in three different positions: $0\,^{\circ}$ to ice deposition, $90\,^{\circ}$ to collect the spectrum, and $180\,^{\circ}$ to electron irradiation. The sketch of the experiment and an overview of the scenario simulated in this work is shown in Fig.~\ref{chamber}. A Nicolet nexus FTIR spectrometer, operating in absorption mode, in the mid-infrared region (4000-800 cm$^{-1}$ or 2.5-12.5 $\mu$m) monitored the chemical changes due to the irradiation of the ice by collecting spectra at a resolution of 2 cm$^{-1}$.

The ice simulating the frozen-plume fallout of Enceladus was made by mixing H$_2$O:CO$_2$:CH$_4$:NH$_3$:CH$_3$OH in several experiments in slightly different proportions. Here we present the results of the 10:1.3:0.4:0.3:0.2 mixture, derived from the column densities (N). The column density (N) of the molecules in the non-irradiated ice was calculated using a modified Lambert-Beer equation (\cite{moll1966}):
\begin{equation}
     N = \frac{ln(10)}{A} \int_{\nu i}^{\nu f} Abs_{\nu}~d_{\nu} \,,
   \end{equation}
where \textit{Abs$_{\nu}$} is the absorbance coefficient integrated at a given frequency interval ($\nu$$_f$ - $\nu$$_i$), and \textit{A} is the band strength value (or \textit{A-value}) for a specific vibrational mode in the $\nu$$_f$ - $\nu$$_i$ region.

According to our results, small changes in the mixture do not affect the spectral results in a measurable way, although a water-poor experiment showed small changes in the formation cross section of some daughter species (data not shown here). This result is discussed below.

The samples were deposited and irradiated normally and isothermally at 20 K with 1 keV electrons, generated by an electron gun (Kimball Physics, EGPS-1011A) with 10 $\mu$A of current in the electron gun power supply. The extraction efficiency of the cathode, according to the manufacturer, is $\sim$90\%; the loss due to the beam spread was calculated using a copper plate coated with phosphorus over the sample holder to measure the area of the electron beam over the area of the ZeSe crystal. The estimated loss is $\sim$30\% which means that the total efficiency is $\sim$60 \% of the current in the power supply. Therefore, considering a penetration depth of 60~$\pm$~10 nm estimated using a Gaussian distribution calculated with the software CASINO (\citealt{hovi1997}), the samples were exposed to an electron beam flux of (3.7$\pm$0.7)$\times$10$^{13}$ electrons cm$^{-2}$ s$^{-1}$, and the final fluence after 210 minutes of irradiation was (4.7$\pm$1)$\times$10$^{17}$ electrons cm$^{-2}$. The deposited energy (dose) was $\sim$2.5 eV molecule$^{-1}$ s$^{-1}$, or a total of $\sim$2.1$\times$10$^{4}$ eV molecule$^{-1}$. Considering the electron flux in Enceladus (adapted from \citealt{para2012}) of $\sim$1.8$\times$10$^{5}$ $\times$ 1 keV electrons cm$^{-2}$ s$^{-1}$, the corresponding Enceladus' time-scale is approximately 8$\times$10$^{4}$ years of exposition to 1 keV electrons from Saturn's magnetosphere and from the solar wind.

During the irradiation, several spectra were collected to monitor changes in the ice. Typically, one spectrum of the ice was collected at every 5, 10, or 30 minutes of irradiation. After irradiation, the sample was heated to room temperature (2 K min$^{-1}$ heating ramp) and a spectrum was collected approximately at every 10 K.

Table~\ref{tab:positions} compares the band positions observed in the IR spectrum recorded in this work with results from the literature and shows the employed literature band strengths (or \textit{A-values}) used to derive the results.

\begin{table*}
\centering
      \caption[]{Band positions of the parental molecules in this experiment in comparison with the values obtained from the literature, and the corresponding band strength (\textit{A}). Comparing results from different references, ice mixtures and temperatures, the estimated uncertainties in the band strengths (\textit{A}) are up to 20\%.}
         \label{tab:positions}
           \begin{tabular} { c c c l c }
            \hline
            \noalign{\smallskip}
            Molecule       & Band position (cm$^{-1}$/ $\mu$m) & Band position (cm$^{-1}$)     & Vibrational mode   &  \textit{A-value}$^{\ast}$ \\
                           & (this work)                       & (literature)$^{\ast}$         &                    &  (cm molecule$^{-1}$)\\
            \noalign{\smallskip}
            \hline
            \noalign{\smallskip}
            H$_2$O             & 1643/ 6.08                    & 1670                        & O-H bend             & 8.4$\times$10$^{-18}$ \\
            CO$_2$             & 2342/ 4.27                    & 2340                        & C=O stretch          & 7.3$\times$10$^{-17}$ \\
            CH$_4$             & 1304/ 7.67                    & 1304                        & C-H deformation      & 6.1$\times$10$^{-18}$ \\
            NH$_3$             & 1124/ 8.89                    & 1115                        & N-H umbrella         & 1.4$\times$10$^{-17}$ \\
            CH$_3$OH           & 1017/ 9.83                    & 1018                        & C-O stretch          & 1.0$\times$10$^{-17}$ \\
            \noalign{\smallskip}
            \hline
            \noalign{\smallskip}
            $^{\ast}$d'Hendecourt \& Allamandola (1986)
         \end{tabular}
    \end{table*}

\section{Results}
\subsection{Infrared spectrum of the pristine ice}
\begin{figure}
     \centering
     \resizebox{\hsize}{!}{\includegraphics{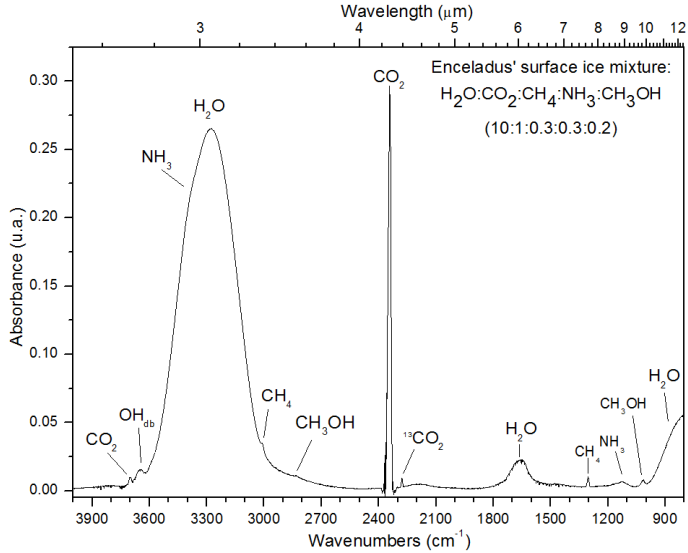}}
     \caption{Spectrum of the pristine (unirradiated) ice (H$_2$O:CO:NH$_3$:CH$_3$OH 10:1.3:0.4:0.3:0.2) at 20 K in the 4000-800 $^{-1}$ (2.5-12.5 $\mu$m region. The most intense bands are identified.}
     \label{pristine}
  \end{figure}

The spectrum of the pristine (unirradiated) ice is shown in Fig.~\ref{pristine}. The most intense bands are the stretching mode of water in 3298 cm$^{-1}$ (3.03 $\mu$m) and the stretching mode of CO$_2$ in 2342 cm$^{-1}$ (4.27 $\mu$m). The weak bands in 3702 and 3630 cm$^{-1}$ are the combination bands (2$\nu$2 + $\nu$3) CO$_2$ and to the dangling bonds of H$_2$O (OHdb), respectively. The very weak shoulder in the 3390 cm$^{-1}$ region is due to the N-H stretch of ammonia, and the shoulder in 3008 cm$^{-1}$ is the C-H stretch of methane. Three methanol bands are distinguishable: the C-H stretch in 2835 cm$^{-1}$, the CH$_3$ deformation in 1479 cm$^{-1}$, and the C-O stretch in 1017 cm$^{-1}$. The strongest methanol band, the O-H stretch, is overlaid by the water O-H stretch in the 3300 cm$^{-1}$ region. Because some O-H stretching bands from water, methanol, and methane, plus the N-H stretch from ammonia, are overlapping in the 3600-3000 cm$^{-1}$ region, no reliable quantitative information can be derived from this region. Therefore the 1643 cm$^{-1}$ (6.08 $\mu$m) band better represents the water in this experiment.

Table~\ref{tab:allbands} shows the most representative bands, their molecular attribution, and the respective vibrational modes as detected in the pristine ice.

\begin{table*}
\centering
      \caption[]{Characterization of all bands identified in the pristine ice spectrum of the simulated Enceladus surface at 20 K. $^\ast$H$_2$O is the dominant, but not the only, molecule in this band.}
         \label{tab:allbands}
           \begin{tabular} { c c c c }
            \hline
            \noalign{\smallskip}
            \multicolumn{2}{c}{Position}                            & Molecule             & Vibrational mode  \\
            cm$^{-1}$       & $\mu$m                                &                      &                   \\
            \noalign{\smallskip}
            \hline
            \noalign{\smallskip}
            3702            & 2.70         & CO$_2$                & comb ($\nu$2 + $\nu$3) \\
            3648            & 2.74         & H$_2$O                & dangling bonds    \\
            3380            & 2.96	       & NH$_3$	               & N-H stretch \\
            3298	        & 3.03         & H$_2$O*	           & O-H stretch \\
            3008	        & 3.32	       & CH$_4$	               & C-H stretch \\
            2835	        & 3.53	       & CH$_3$OH	           & C-H stretch \\
            2342	        & 4.27	       & CO$_2$	               & C=O stretch \\
            2278	        & 4.39	       & $^{13}$CO$_2$	       & C=O stretch \\
            1643	        & 6.09	       & H$_2$O	               & O-H bend \\
            1479	        & 6.76	       & CH$_3$OH	           & CH$_3$ deformation \\
            1304	        & 7.67	       & CH$_4$	               & C-H deformation \\
            1124	        & 8.90	       & NH$_3$	               & NH umbrella \\
            1017	        & 9.83	       & CH$_3$OH	           & C-O stretch \\
            ~800	        & 12.5	       & H$_2$O	               & libration \\
            \noalign{\smallskip}
            \hline
         \end{tabular}
    \end{table*}


\subsection{Ice processing by 1 keV electrons}

Table~\ref{tab:coldens} shows the initial and final column density after electron irradiation of the parent molecules. The column densities were derived from the peak area in the IR spectrum using Eq. (1). The normalized variation of the column density over the fluence is shown in Fig.~\ref{desrate}. H$_2$O has lower destruction ratio. This may be because the radicals produced by the H$_2$O dissociation can react quickly , recycling water in the process. OH$^-$ radicals from the dissociation of methanol can easily react with H$^+$ to produce water. In addition, some irradiation products have vibrational modes in the same region as the water bending mode, which may increase the absorbance signal in this region. On the other hand, NH$_3$ has the highest destruction ratio of all parental molecules. Data from the Cassini mission show that ammonia is very low abundant in the Enceladus plume (\citealt{spen2006}).

\begin{table*}
\centering
      \caption[]{Variation in the column density of the parental molecules due to electron irradiation. The band strength values used to calculate the column densities are shown in Table~\ref{tab:allbands}.}
         \label{tab:coldens}
           \begin{tabular} { c c c c c }
           \hline
            \noalign{\smallskip}
            Molecule       & Initial column density (N$_i$)             & Final column density (N$_f$)           & Destroyed molecules [\%]  \\
                             & (molecule cm$^{-2}$)                     & (molecule cm$^{-2})$                   &  \\
            \noalign{\smallskip}
            \hline
            \noalign{\smallskip}
            H$_2$O          & (40~$\pm$~7$)\times$10$^{16}$           & (32~$\pm$~6$)\times$10$^{16}$       & 20 \\
            CO$_2$          & (5.4~$\pm$~1.0$)\times$10$^{16}$          & (3.7~$\pm$~0.7$)\times$10$^{16}$      & 31 \\
            CH$_4$          & (1.7~$\pm$~0.3$)\times$10$^{16}$          & (1.1~$\pm$~0.2$)\times$10$^{16}$      & 35 \\
            CH$_3$OH        & (0.9~$\pm$~0.2$)\times$10$^{16}$          & (0.5~$\pm$~0.1$)\times$10$^{16}$      & 44 \\
            NH$_3$          & (1.3~$\pm$~0.2$)\times$10$^{16}$          & (0.5~$\pm$~0.1$)\times$10$^{16}$      & 61 \\
            \noalign{\smallskip}
            \hline
         \end{tabular}
    \end{table*}

\begin{figure}
     \centering
     \resizebox{\hsize}{!}{\includegraphics{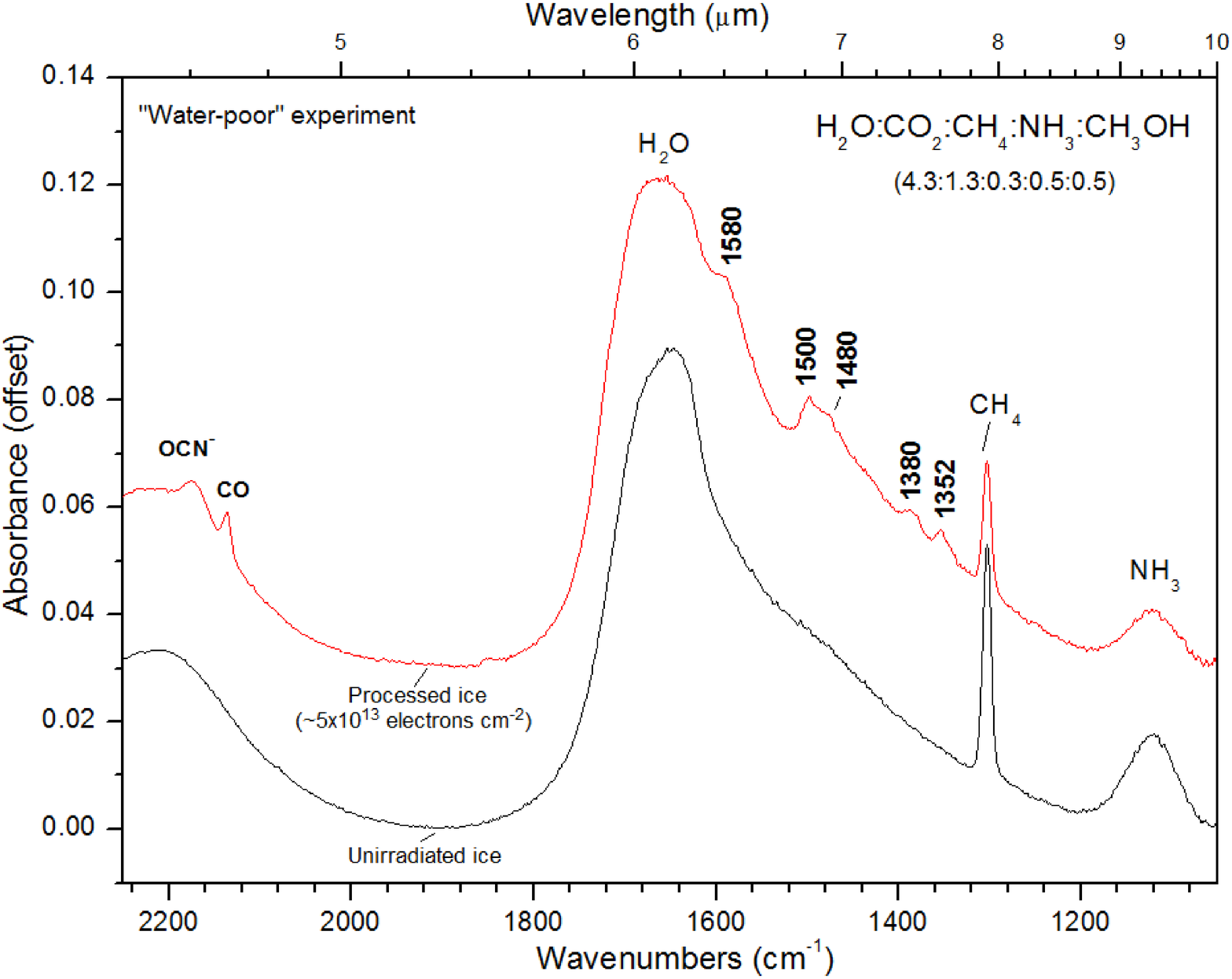}}
     \caption{Evolution of the water-poor ice after 210 minutes of irradiation by 1 keV electron isothermally at 20 K. The final fluence was approximately 4.7$\times$10$^{13}$ electrons cm$^{-2}$. The new bands  due to the processing of the ice are identical in both experiments.}
     \label{newbands}
  \end{figure}

Fig.~\ref{newbands} shows the IR spectrum of the water-poor experiment before and after the bombardment by 1 keV electrons. The new bands in the ice are identical both in the Enceladus simulation ice (experiment 1) and the water-poor ice (experiment 2). Experiment 1 represents our best attempt do create an ice analogue to plume fallout in cold regions of Enceladus (e.g. north pole), and experiment 2 was made to compare the possible effects from the reducing of the amount of water in the ice matrix. The region of the infrared spectrum inside the 2250 - 1000 cm$^{-1}$ (4.44 - 10 $\mu$m) range shows more daughter species. The new bands are identified in bold by their positions in cm$^{-1}$. The evolution of the new bands was monitored using the software \texttt{OMNIC}. The band in the 2170 cm$^{-1}$ (4.60 $\mu$m) region is a common product from the processing of C, O, and NH$_3$ containing ices, and we assign this band to the cyanate anion (OCN$^-$). The OCN$^-$ is not a first-generation radical (\citealt{moor1983}), and the intermediary products of the OCN$^-$ reaction route are good indicators of the molecules produced in the processed ice. Some reaction routes that may lead to the formation of the cyanate anion are indicated below (adapted from \citealt{vanb2004}; \citealt{pill2010b}; \citealt{jone2011}). Molecules in boldface indicate a specie detected in our experiments.

\begin{equation}
  \textbf{CO$_2$}  \stackrel{e^-}{\longrightarrow} \textbf{CO} + O \\
\end{equation}
\begin{equation}
  \textbf{NH$_3$}  \stackrel{e^-}{\longrightarrow} NH_2 + H \\
\end{equation}
\begin{equation}
  \textbf{NH$_3$} + NH_3^+ \longrightarrow \textbf{NH$_4^+$} + NH_2\\
\end{equation}
\begin{equation}
  NH_2 + \textbf{CO} \longrightarrow CONH_2 \\
\end{equation}
\begin{equation}
  CONH_2  \longleftrightarrow HNCO + H  \\
\end{equation}
\begin{equation}
  HNCO + \textbf{H$_2$O} \stackrel{e^-}{\longrightarrow} \textbf{OCN$^-$} + H_3O^+ \\
\end{equation}
\begin{equation}
  HNCO + \textbf{NH$_3$}  \longrightarrow \textbf{OCN$^-$} + \textbf{NH$_4^+$} \\
\end{equation}

A CO band was identified in the infrared spectrum in the 2140 cm$^{-1}$ (4.67 $\mu$m) region. Traces of CO are visible in the IR spectrum after just a few minutes of irradiation (data not shown here), and it remains one of the most visible new bands in the experiment. The possible CO formation pathway is described in reaction (2).

The new band in 1500 cm$^{-1}$ (6.66 $\mu$m) was also observed in similar experiments (e.g. \citealt{dhena1986}; \citealt{bern1995}; \citealt{ober2009}; \citealt{gera1996}). This band could be assigned to formaldehyde (H$_2$CO) if a band from the CH$_2$ scissoring were present in 1720 cm$^{-1}$, but this region is dominated by water, and is not possible to confirm any other bands in this region of the spectrum. Studies by \cite{pill2010a} showed a band in 1500 cm$^{-1}$ with a profile very similar to the same band in our experiments. This band was assigned to the NH$_4^+$ ion. The production of NH$_4^+$ may be explained by reactions (4) and (8).

The 1383 cm$^{-1}$ (7.23 $\mu$m) band was detected by \cite{jhee2013} through electron irradiation of NH$_3$:CH$_3$OH ices and by \cite{vanb2004} through UV photolysis of H$_2$O:CO:NH$_3$ ices. Both authors assigned this band to the production formamide (HCONH$_2$). \cite{vanb2004} highlighted the importance of HCONH$_2$ as a precursor of OCN$^-$ in interstellar ice grains and in star forming regions. The weak shoulder in the 1580 cm$^{-1}$ (6.33 $\mu$m) band was also detected by \cite{bern1995} and was assigned to an H$-$N$-$H scissoring of amides, such as formamide. This band reinforces the possibility that HCONH$_2$ might be produced in our experiment. Considering previous reactions (2), (3), (4) and (5), the suggested reaction pathway to the production of formamide is (adapted from \cite{jone2011}):
\begin{equation}
  \textbf{NH$_3$} + \textbf{CO} \longrightarrow \textbf{HCONH$_2$} \\
\end{equation}
\begin{equation}
  \textbf{CO} + H \longrightarrow HCO \\
\end{equation}
\begin{equation}
  HCO + NH_2 \longrightarrow \textbf{HCONH$_2$} \\
\end{equation}
\begin{equation}
  CONH_2 + H \longrightarrow \textbf{HCONH$_2$}. \\
\end{equation}
 CH$_2$OH was detected by \cite{gera1996} through UV irradiation of pure methanol (CH$_3$OH), and by \cite{jhee2013} through electron irradiation of a methanol-containing ice. In both cases, the CH$_2$OH band was identified in the 1352 cm$^{-1}$ (7.39 $\mu$m) region, the same position as in this work. We suggest some simple formation routes for the CH$_2$OH:
\begin{equation}
  \textbf{H$_2$O} \stackrel{e^-}{\longrightarrow} OH + H \\
\end{equation}
\begin{equation}
  CH_3OH + H \longrightarrow \textbf{CH$_2$OH} + H_2 \\
\end{equation}
\begin{equation}
  CH_3OH  + OH \stackrel{e^-}{\longrightarrow} \textbf{CH$_2$OH} + \textbf{H$_2$O}. \\
\end{equation}
Table~\ref{tab:newspecies} summarizes the main products detected through electron irradiation of the samples and their tentative assignments. It is noteworthy that the bands in 1383 cm$^{-1}$ and 1250 cm$^{-1}$ were also detected by \cite{muno2003} and \cite{bern1995} through the processing of water, CO, NH$_3$, and methanol-containing ices, and these bands were assigned to HMT (hexamethylenetetramine - C$_6$H$_{12}$N$_4$). The masses described in the last column of Table~\ref{tab:newspecies} are compared with the masses detected in Enceladus' plume by the Cassini mission. Originally, some of these masses were assigned to different species (e.g. \citealt{wait2009}). Our infrared spectroscopy results can be interpreted as a complement to the results of the Cassini's mass spectrometer, demonstrating that species not yet detected in Enceladus because of instrumental limitations may be present in that rich environment. These results are discussed below.
\begin{table*}
\centering
      \caption[]{Assignment of the main bands produced after 210 minutes of electron beam irradiation (fluence of 4.7~$\times$10$^{17}$ electrons cm$^{-2}$).}
         \label{tab:newspecies}
           \begin{tabular} { c l c l }
            \hline
            \noalign{\smallskip}
            Band position       & Tentative                    & Molecular mass               & Assignment reference              \\
            (cm$^{-1}$)         & Assignment                   & (amu)                        &         \\
            \noalign{\smallskip}
            \hline
            \noalign{\smallskip}
            2170                & OCN$^-$                      & 42                     & \cite{moor1983}                      \\
            2140                & CO                           & 28                     &                                      \\
            1580                & -COO$^-$                     & 44                     & \cite{muno2003}                      \\
            1500                & NH$_4^+$ and/or H$_2$CO      & 18/ 30                 & \cite{gera2004} / \cite{schu1993}    \\
            1480                & NH$_4^+$ and/or C$_m$H$_n$   & 18/ 24-60              & \cite{gera2004} / \cite{moor1983}    \\
            1383                & HCONH$_2$                    & 45                     & \cite{gera2004}                      \\
            1352                & CH$_2$OH and/or CH$_3$CHO    & 31/ 44                 & \cite{schu1993}/ \cite{moor1998}     \\
            1250                & HMT (?)                      & n/a                    & \cite{bern1995}                      \\
            \noalign{\smallskip}
            \hline
         \end{tabular}
    \end{table*}

\subsection{Destruction and formation cross-sections of molecular species in the ice}

The evolution of the area of selected peak in the in the infrared spectra as a function of the fluence allows us to quantify the destruction or the formation cross-section of a given molecular species.
We used the following equation to best fit this area dependence with the fluence:
\begin{equation}
        A-A_0 = A_{\infty} \cdot (1 - e^{(-\sigma_{d,f} \cdot F)})\,, \\
\end{equation}
where \textit{A}, \textit{A$_0$} and \textit{A$_{\infty}$} are the selected area of the infrared band related with a specific molecule at a given fluence (\textit{A}), at the beginning of the experiment (\textit{A$_0$}, unirradiated sample), and at the highest fluence (\textit{A$_{\infty}$},  the terminal  fluence). In  this  expression,  $\sigma$  represents  the  formation  cross-section  ($\sigma_f$)  for  new  species, or the destruction cross-section ($\sigma_d$) for parental species, depending on each case. The cross-sections are expressed in units of cm$^{2}$. \textit{F} indicates the irradiation fluence (in units of  electrons cm$^{-2}$). This single equation is a modified version of the equation commonly employed in the literature for similar experiments (e.g. \citealt{pill2014a}; \citealt{alme2014}). The advantage of this methodology is that the chemical changes in both parent and daughters species caused by sample processing can be quantified by the same expression. Moreover, since expression \textit{(A-A$_0$)/A$_{\infty}$} varies between 0 and 1, this equation normalizes the data. This equation was also recently employed by \cite{pill2014b} to quantify the changes in astrochemical ices caused by UV and X-ray photons.

\begin{figure}
     \centering
     \resizebox{\hsize}{!}{\includegraphics{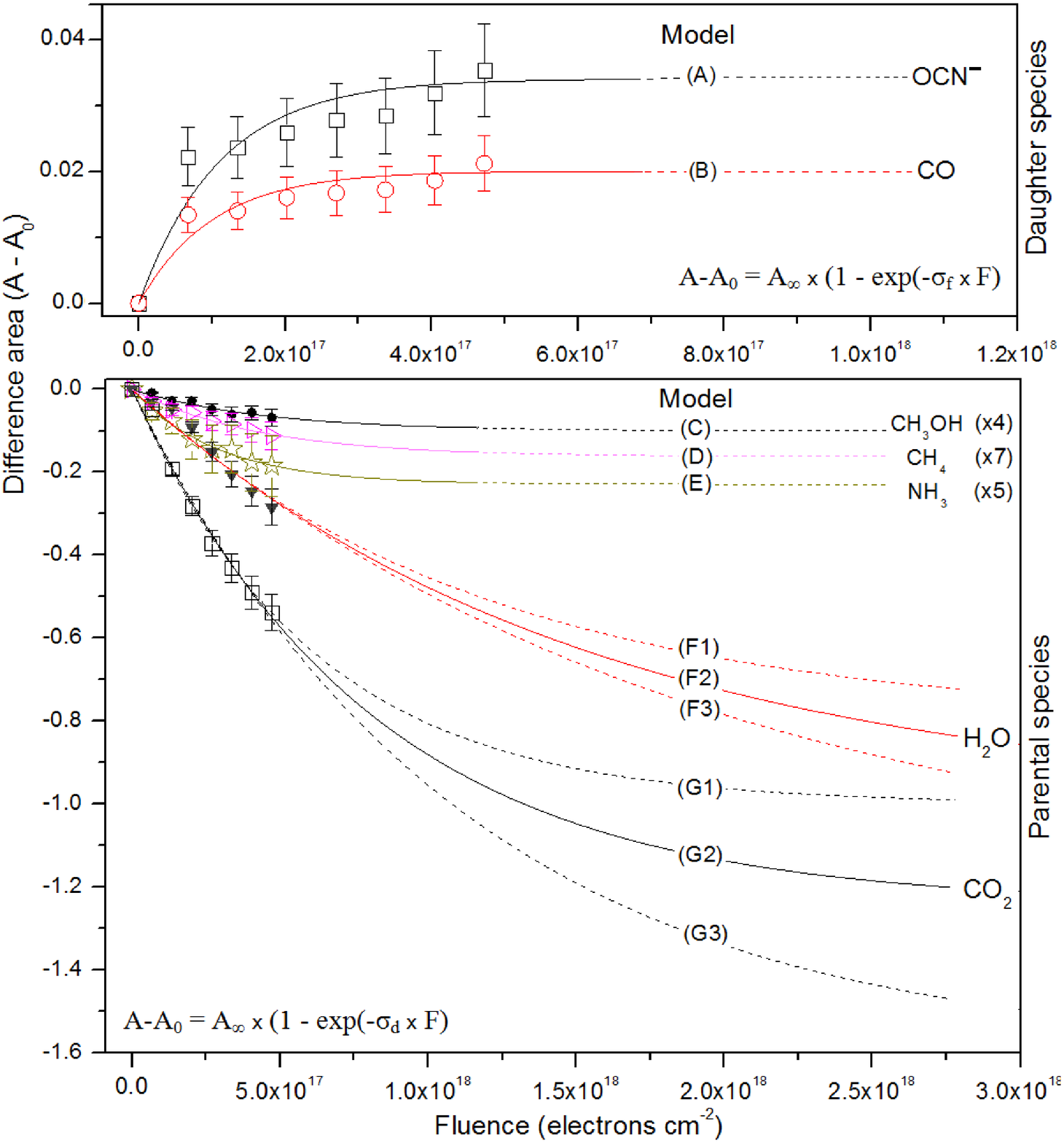}}
     \caption{Difference area over the fluence. The upper panel shows the fit for the two most abundant daughter species, the bottom panel shows the fit for selected bands of parental species. The points are the experimentally obtained values and the lines are the best fit using Equation (17). The difference areas of CH$_3$OH, CH$_4$, and NH$_3$ were multiplied by a constant, as indicated, for better visualization. However, the values of A$_{\infty}$ and the cross-section of these species were calculated from the original data. The results from models F\textit{n} and G\textit{n} are shown in Table~\ref{tab:crosssections}. In this figure, higher amplitude does not mean higher destruction or formation of the molecules.}
     \label{crosssections}
  \end{figure}

\begin{figure}
     \centering
     \resizebox{\hsize}{!}{\includegraphics{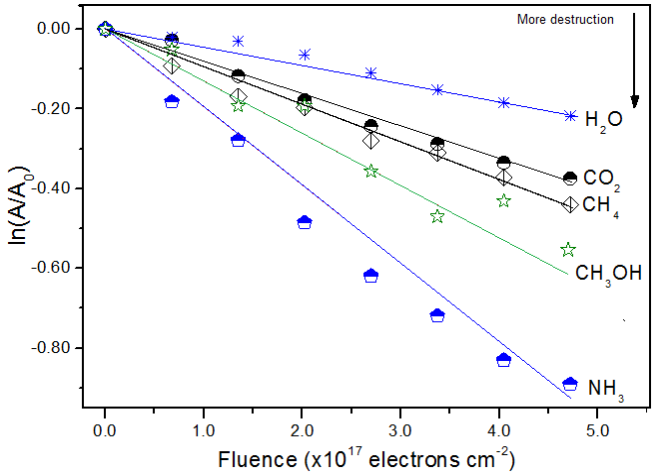}}
     \caption{Destruction rate of the parental species. As shown in Table 3, H$_2$O and NH$_3$ are the less and most destroyed molecules, respectively. The points are the experimentally obtained values. The lines are a guide for the eyes.}
     \label{desrate}
  \end{figure}

Equation (16) presumes that daughter species are not significantly destroyed or sputtered by radiation, and that parent species are not significantly produced from daughter species. This hypothesis is incorrect for experiments using heavy and swift ions (e.g. \citealt{barr2011}; \citealt{berg2014}). In the current experiment, the main uncertainty in determining the cross-section is the determining the correct A$_{\infty}$ in Eq. (16). As seen in Fig.~\ref{crosssections}, we estimated that the actual value of A$_{\infty}$ would be reached in fluences of approximately $1\times10^{18}$ electrons cm$^{-2}$, the equivalent of $\sim$7 hours of irradiation, for CH$_4$, NH$_3$, and CH$_3$OH, and $3\times10^{18}$ electrons cm$^{-2}$, equivalent to $\sim$22 hours of irradiation, for CO$_2$ and H$_2$O. Because of the doubt determining the value of A$_{\infty}$ for H$_2$O and CO$_2$, three models were calculated for each one of these molecules. All the calculated cross-sections are listed in Table~\ref{tab:crosssections}, including the different models used for H$_2$O and CO$_2$. The error arising from the use of different models was determined to be approximately 30\%. In Fig.~\ref{crosssections}, positive values indicate production (daughter species) and negative values indicate destruction (parental species). Note that the amplitude of the difference area in Fig.~\ref{crosssections} does not mean a higher destruction or production rate of the molecules. The destruction rate of parental molecules is shown in Fig.~\ref{desrate}.

\begin{table*}
\centering
      \caption[]{Cross-sections (destruction or formation) of the molecules in the ice. The error in the determination of these parameters is estimated to be up to 30 \%. The cross-sections are expressed in units of cm$^2$.}
         \label{tab:crosssections}
           \begin{tabular} { l l l l l }
           \hline
           \noalign{\smallskip}
           \multirow{2}{*}{Molecule}             & \multirow{2}{*}{Position (cm$^{-1}$)}   & \multirow{2}{*}{Model} & \multirow{2}{*}{A$_\infty$}   & Cross section ($\sigma_{d,f}$)   \\
                                &                        &       &              & (formation or destruction)       \\
           \noalign{\smallskip}
           \hline
           \noalign{\smallskip}
            Daughter species  & & & & \\
           \noalign{\smallskip}
           \hline
           \noalign{\smallskip}
              OCN$^-$              & 2170       & (A)     &  0.03     &(9.0$\pm$0.3$)\times10^{-18}$  \\
              CO                   & 2140       & (B)     &  0.02     &(9.9$\pm$0.3$)\times10^{-18}$  \\
           \noalign{\smallskip}
           \hline
           \noalign{\smallskip}
            Parental species  & & & & \\
           \noalign{\smallskip}
           \hline
           \noalign{\smallskip}
            CH$_3$OH                 & 1017           & (C)     & -0.03    &(1.4~$\pm$~0.4$)\times10^{-18}$      \\
            CH$_4$                   & 1304           & (D)     & -0.02    &(2.3~$\pm$~0.8$)\times10^{-18}$      \\
            NH$_3$                   & 1124           & (E)     & -0.05    &(3.4~$\pm$~1$)\times10^{-18}$        \\
            \hline
            \multirow{3}{*}{H$_2$O}  &                & (F1)    & -0.8     &(8~$\pm$~3$)\times10^{-19}$          \\
                                     & 1643           & (F2)    & -1.0     &(6~$\pm$~2$)\times10^{-19}$          \\
                                     &                & (F3)    & -1.5     &(4~$\pm$~1$)\times10^{-19}$          \\
            \hline
            \multirow{3}{*}{CO$_2$}  &                & (G1)    & -1.0     &(1.6~$\pm$~0.5$)\times10^{-18}$      \\
                                     & 2342           & (G2)    & -1.2     &(1.2~$\pm$~0.4$)\times10^{-18}$      \\
                                     &                & (G3)    & -1.6     &(9~$\pm$~3$)\times10^{-19}$          \\
           \noalign{\smallskip}
           \hline
         \end{tabular}
    \end{table*}

The values of the half-life in the laboratory can be determined by the expression (\citealt{pill2014a})
\begin{equation}
  \tau_{1/2} (lab) =\frac{ln(2)}{k}, \\
\end{equation}

were \textit{k} is the electron-dissociation rate in units of s$^{-1}$, calculated as a product of the destruction cross-section by the electron beam flux. The values of \textit{k} and $\tau_{1/2}$ are listed in Table~\ref{tab:halflife}.

\begin{table}
\centering
      \caption[]{Dissociation rate and laboratorial half-life for parental species, calculated for a flux of $3.7\times10^{13}$ electrons s$^{-1}$, using the mean value of the cross-sections from Table~\ref{tab:crosssections}.}
         \label{tab:halflife}
           \begin{tabular} { l l l }
           \hline
           \noalign{\smallskip}
            Molecule            & k$_{lab}$ ($s^{-1}$)    & $\tau_{1/2}$ (s) \\
           \noalign{\smallskip}
           \hline
           \noalign{\smallskip}
            CH$_3$OH            & $5.3\times10^{-5}$    & $1.3\times10^{4}$ \\
            CH$_4$              & $8.7\times10^{-5}$    & $7.9\times10^{3}$ \\
            NH$_3$              & $1.3\times10^{-4}$    & $5.4\times10^{3}$ \\
            H$_2$O              & $2.2\times10^{-5}$    & $3\times10^{4}$ \\
            CO$_2$              & $4.5\times10^{-5}$    & $1.5\times10^{4}$ \\
           \noalign{\smallskip}
           \hline
         \end{tabular}
    \end{table}

\begin{figure}
     \centering
     \resizebox{\hsize}{!}{\includegraphics{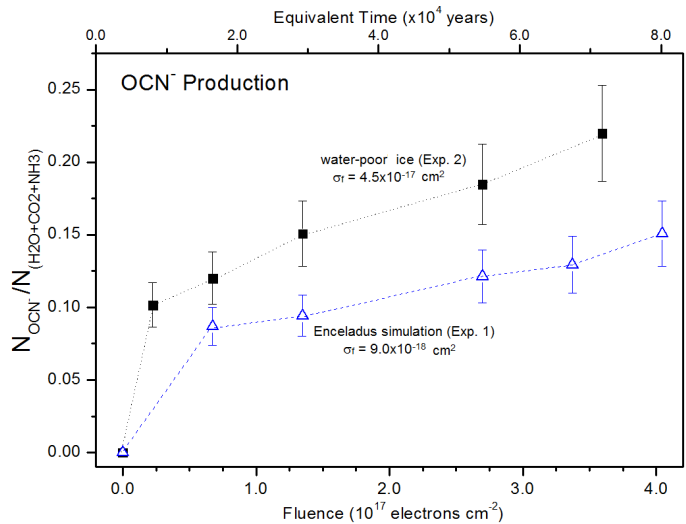}}
     \caption{Comparison of the OCN$^-$ column density divided by the sum of the of H$_2$O, CO$_2$, and NH$_3$ in water-poor experiment (black square) and water-rich experiment (blue triangle), both irradiated at 20 K. The lines are a guide for the eyes. The y-axis is multiplied by 100.}
     \label{OCN}
  \end{figure}

The IR spectrum from the water-poor experiment (4.3:1.3:0.3:0.5:0.5) shows no significant changes in terms of the produced species (i.e. the qualitative result is very similar), but the comparison of the formation cross-section of the OCN$^-$ reveals a dependence on the NH$_3$ and CO$_2$ abundance in the water matrix. Comparing the water-rich and the water-poor experiments, the formation cross-section of the OCN$^-$ is 3.5 times higher in the water-poor experiment. In this sense, the OCN$^-$ abundance in frozen bodies in the Solar system (such as comets, asteroids, and moons), in star-forming regions (such as W33A, RAFGL 7009S), and in interstellar ices may give a good indication of the ratio of molecules such as H$_2$O, CO$_2$, and NH$_3$ in the history of that object. Fig.~\ref{OCN} shows a comparison of the column density of the OCN$^-$ divided by the sum of the column densities of H$_2$O, CO$_2$, and NH$_3$  (N$_{(OCN^-)}$/ N$_{(H_2O + CO_2 + NH_3)}$) in both experiments. The water-poor ice was made by decreasing the water pressure in the mixture chamber.

\begin{figure}
     \centering
    \resizebox{\hsize}{!}{\includegraphics{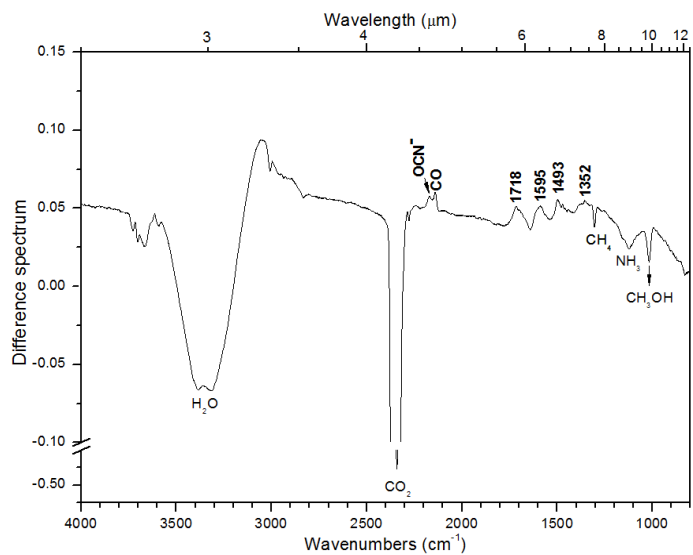}}
     \caption{Difference spectrum (irradiated at final fluence minus pristine ice). The new species are highlighted in bold.}
     \label{difference}
  \end{figure}

Because of the great number of overlaps in the vibrational modes in such a complex mixture, the investigation of the new bands that appear upon electron irradiation can be concluded with the difference spectra (as discussed by \citealt{dhena1986}). The difference is taken between the spectrum of the non-irradiated and that of the irradiated ice. In the difference spectrum, the bands that grow upon electron irradiation point upward, those that diminish point downward. Fig.~\ref{difference} shows the result of subtracting of the irradiated ice (fluence $4.7\times10^{17}$ electrons cm$^{-2}$) by the pristine ice (unirradiated). The daughter species are highlighted in bold.

The subtraction of two spectra may artificially enhance weak bands or even show the result of contamination instead of irradiation-produced molecules. Thus, the results from the subtraction spectrum should be taken with great caution. Nonetheless, Fig.~\ref{difference} shows some distinct bands, especially in the 1750-1500 cm$^{-1}$ interval, a region dominated by water, bypassing (to some extent) the overlap bands problem in this region.

The 2140 cm$^{-1}$ band in the difference spectrum is due to the CO molecule. The 1718 cm$^{-1}$ might be due to the CH$_2$ scissoring of formaldehyde (H$_2$CO). A band assigned to formamide (HCONH$_2$) was detected by \cite{vanb2004} in 1352 cm$^{-1}$ wavenumbers in similar experiments. The subtraction spectrum reinforces the hypothesis of the detection of CO and OCN$^-$, and reveals the possibility of the production of -COO$^-$ (in carboxylic acids) and formamide (HCONH$_2$) in our experiment. Although the OCN$^-$ was not detected by the Cassini instruments, our results shows that this molecule is probably produced in some regions of Enceladus.


\section{Discussion}

Quantifying the electron-induced ice chemistry in such complex mixtures using only FTIR spectroscopy is a difficult task, since there is a multitude of irradiation-products with overlapping features. Nevertheless, this mixture can be considered a good approximation of the ice in resurfaced and cold regions of Enceladus, and is worth to be studied. In this experiment, the molecules detected through electron irradiation are probably just a small fraction of the molecules produced in a chemically rich environment such as Enceladus and, by extension, in several other frozen objects in the solar system. In this sense, the reaction routes (see the Results section) of the detected species are a good indication of the chemical enrichment and the increase of complexity in such environments. Some detected molecules, such as OCN$^-$ and HCONH$_2$, require multiple reaction steps to be produced, and although the intermediary species were not abundant enough to be detected in our experiments, it is certainly being produced. The study of the molecules in binary mixtures may give clearer results in terms of the new species (e.g. \citealt{jhee2013}; \citealt{dhen1985}; \citealt{dhena1986}; \citealt{dhenb1986}; \citealt{muno2003}; \citealt{bern1995}), but it may not be a realistic representation of the physico-chemical processes that occur in complex ice bodies such as Enceladus.

So far, the most reliable and detailed data about the chemistry in Enceladus was provided by the instruments onboard the Cassini spacecraft, especially during the flybys (\citealt{wait2006} and \citealt{wait2009}). The INMS instrument detects molecules with masses ranging from 1 to 80 amu in the plume, but it does not differentiate between two or more species with same molecular mass. For example, our results show that the mass 30 amu, originally assigned to N$_2$ and/or C$_2$ organics by \cite{wait2009}, may also be assigned to formaldehyde (H$_2$CO). Mass 31 amu may be assigned to CH$_2$OH and the mass 45 amu to formamide (HCONH$_2$). Higher masses (>50 amu), assigned to molecules such as C$_4$H$_n$ and C$_6$H$_n$, might be caused by fragments of simple amino acids, such as glycine and alanine. Glycine was tentatively detected by \cite{ober2009} in a UV photolysis of similar ices.

Ammonia was not detected in Enceladus in the firsts flybys, probably because of the low abundance of only 0.8\% in volume with respect to water (\citealt{wait2009}). In our experiments, NH$_3$ is the molecule most destroyed by irradiation, and conversely, H$_2$O is the least destroyed molecule. Not coincidentally, CO$_2$ presents the second-lowest destruction ratio in our experiments, and it is the second-most abundant molecule in Enceladus (except for H$_2$) according to data from Cassini. Therefore the ratio of the molecules in the surface of a frozen body is strongly related with physicochemical processes triggered by ionizing radiation (including photolysis and radiolysis, processes not studied in this work). In other words, the current chemical composition of a frozen object in the Solar system depends not only on its original chemistry, but is strongly related with the processing of the ice. Therefore, laboratory experiments simulating the interaction of ionizing radiation with icy surfaces at ultralow temperatures offer a remarkable opportunity to understand the chemical processing of primitive objects and allow for an common stream of the chemical time-line dating back to the origin of the Solar system.


\section{Conclusions}
We simulated some physico-chemical processes that probably occur in cold regions of Enceladus, such as the north pole. Although simulations using the software CASINO showed that 1 keV electrons only penetrate the surface up to 60 nm, other ionizing agents, such as X-rays and, especially, cosmic rays, may penetrate several times deeper than this, resulting in even more complex physico-chemical effects. The influence of the temperature on the chemical transformation of the ice in environments such as those of Enceladus (the temperature gradient of Enceladus ranges from 30 K to more than 160 K), as well as of the ionizing agent (e.g. UV photons, protons, energetic electrons, X-rays, and cosmic rays), must be investigated using complementary diagnostic techniques, such as laser-induced time-of-flight mass spectroscopy (Bergantini et al. in preparation).

Our experiments have shown the production of molecules not yet detected in Enceladus, such as NH$_4^+$, HCONH$_2$, CH$_2$OH, and HMT, from the processing of an analogue ice. The fact that our experiments were carried out in lower temperatures in comparison with Enceladus means that some produced molecules, such as CO, will be more efficiently trapped in the ice than in the hotter regions of Enceladus, making the detection of these molecules in the IR spectrum more evident. Nonetheless, there are probably no qualitative difference in terms of the produced species in our experiments when compared with the processes that occur on Enceladus. The literature shows that some species are preferably produced in methanol-containing ices. In this sense, the relative abundance of methanol in Enceladus can be inferred from possible daughter species, such as H$_2$CO and HMT, and vice versa. The study also showed that there is a strong correlation between the destruction cross-sections of the parental species in our results and the abundance of molecules detected in Enceladus by the Cassini mission, with a low destruction for H$_2$O and CO$_2$ and a high destruction for NH$_3$ and CH$_3$OH, which indicates that the molecular abundances in a frozen body of the Solar system may depend heavily on the processes triggered by ionizing radiation.

Astronomical observations show that the base chemistry in Enceladus (H$_2$O, CO$_2$, NH$_3$, CH$_4$, plus other molecules in minor abundance) is similar to ice-covered grains in protostellar disks (e.g. \citealt{herb1973}; \citealt{muno2002}) and it is also very similar to the composition of cometary comae (\citealt{wait2009}; \citealt{bock2004}). This is probably related with the early stages in the formation of the Solar system, when the gas and dust from the protostellar cloud agglomerate to form lager structures, such as planetesimals.  If the protostellar ice is not destroyed at this stage, minor bodies like comets may deposit these molecules on the surface of planets and moons of the planetary system (\citealt{viss2009}). In this context, the study of Enceladus may help to better understand the early stages of the Solar system in terms of its chemical evolution and prebiotic chemistry.


\begin{acknowledgements}
The author would like to acknowledge the Department of Physical Sciences of The Open University and the FAPESP agency. This work was supported by the FAPESP agency under projects No. 2012/17248-2 and 2009/18304-0.
\end{acknowledgements}

\bibliographystyle{aa} 


\end{document}